# Tunable spin reorientation transition and magnetocaloric effect in $Sm_{0.7-x}La_xSr_{0.3}MnO_3$ series

M. Aparnadevi and R. Mahendiran[1]

Department of physics, 2 Science Drive 3, Faculty of Science,

National University of Singapore, Singapore – 117542, Singapore

Abstract

We report electrical resistivity, magnetic and magnetocaloric properties in $Sm_{0.7-x}La_xSr_{0.3}MnO_3$ series for x= 0, 0.1, 0.2, 0.3, 0.4, 0.5, 0.6, 0.65, and 0.7. All the compounds show second order paramagnetic to ferromagnetic transition at $T = T_c$ which is tunable anywhere between 83 K and 373 K with a proper choice of the doping level (x). The insulating ferromagnet x= 0 transforms to a ferromagnetic metal below $T_c$ for x= 0.1 and the insulator-metal transition temperature shifts up with increasing x. The magnetization (M) exhibits an interesting behavior as a function of temperature and doping level. The field-cooled $M(T)$ of all but x= 0.7 compounds show a cusp at a temperature $T^*$ much below $T_c$. While the $T_c$ increases monotonically with increasing x, $T^*$ increases gradually, attains a maximum value ($T^* = 137$ K) for x= 0.6 and decreases rapidly thereafter. It is suggested that the decrease of $M(T)$ below $T^*$ is due to ferrimagnetic interaction between Sm(4f) and Mn(3d) sublattices that promotes spin-reorientation transition of the Mn-sublattice. The observed anomalous feature in $M(T)$ does not have impact on the dc resistivity. Magnetic entropy change ($\Delta S_m$) was estimated from magnetization isotherms. The sign of $\Delta S_m$ is found to change from negative above $T^*$ to positive below $T^*$ indicating the coexistence of normal and inverse magnetocaloric effects. $\Delta S_m$ is nearly composition independent ($-\Delta S_m= 1.2\pm0.2$ J/Kg K for $\Delta H= 1$ Tesla) and refrigeration capacity lies between 40 and 50 J/kg K for $0.1\leq x\leq 0.6$. We show scaling of magnetic entropy change under different magnetic fields and analysis of critical exponents associated with the phase transition in x= 0.6 compound. The tunability of Curie temperature with nearly constant $\Delta S_m$ value along with high refrigeration capacity makes this series of compounds interesting for magnetic refrigeration over a wide temperature range.

---

[1] Author for correspondence (phyrm@nus.edu.sg)

**Introduction**

The discovery of giant magnetic entropy change associated with the first-order magneto-structural transition in $Gd_5Si_2Ge_2$ rekindled interest in new magnetocaloric materials that can be used as solid state refrigerants in magnetic refrigeration.[1] Magnetic refrigeration is an emerging green and energy efficient technology that can outperform the conventional vapor-compression refrigeration technique. Recently, large magnetocaloric effect near room temperature has been reported in other compounds showing first-order magnetic phase transition such as $Mn(Fe,P)As$,[2] $La(Fe,Si)_{13}$,[3] $MnFe(P,Si,Ge)$ etc.[4] However, high cost, large hysteresis found in thermal and field sweep cycles and mechanical instability due to cracks developed in the first order transition pose certain challenges that have to be overcome before these materials can find applications. A magnetocaloric material is characterized by three important quantities: (1) Magnetic entropy change ($\Delta S_m$), (2) Adiabatic temperature change ($\Delta T_{ad}$), and (3) refrigerant capacity $RC = \int_{T_1}^{T_2} \Delta S_m(T) dT$. The *RC* value represents the amount of heat that can be transferred between a cold reservoir at temperature $T_1$ and hot reservoir at temperature $T_2$ in one ideal refrigeration cycle. Besides large $\Delta S_m$ and $\Delta T_{ad}$, a large *RC* value is highly preferred by practical applications.

Although $\Delta S_m$ in a second order transition is smaller than that in the first-order transition, *RC* can be higher in the former case because of the broadness of the $\Delta S_m$ peak in temperature scale. Many materials have been explored in recent years with a primary aim of optimizing MCE for application in magnetic refrigeration at room temperature. However, magnetic refrigeration over a wide temperature (1-300 K) is also required since the current state of art adiabatic demagnetization refrigerators (ADR) make use of paramagnetic salts which show larger magnetocaloric effect below 1 K. Hence, magnetic materials exhibiting large MCE at low temperatures are also of interest. Since the magnetocaloric effect is dominant around the Curie temperature, magnetic refrigeration over a wide temperature requires stack of materials with varying $T_c$. Full realization of magnetic refrigeration technology relies on finding materials with attractive MCE in a magnetic field of $H$= 1-2 T that can be produced by Nd-Fe-B based permanent magnets.

Perovskite manganites of the general formula $R_{1-x}A_xMnO_3$ (R= $La^{3+}$, $Nd^{3+}$, etc. and A= $Ca^{2+}$, $Sr^{2+}$ etc.) are also considered to be one of the potential materials for magnetic refrigeration.[5,6,7,8,9] Very recently, C. Bahl *et al*. showed that a polycrystalline slab of $La_{0.7}Ca_{0.3}Sr_{0.3}MnO_3$ decreases temperature as much as 9 K when exposed to a magnetic field of 1 T.[10] Perovskite manganites are cheaper than the intermetallic alloys and are easier to prepare. Another advantage of manganite is that their magnetic phase transition temperature can be tuned anywhere between 80 K and 400 K with proper hole content (x) and stacks of materials with varying $T_c$ can be prepared by simple and low-cost materials processing technique such as tape casting.[10] The end members x= 0 and 0.7 of $Sm_{0.7-x}La_xSr_{0.3}MnO_3$ are ferromagnets with $T_c$= 83 K and 372 K, respectively.[11] Hence, $T_c$ in this series is tunable over a wide temperature with La content. There is also another motivation to investigate this series. $SmMnO_3$ is a canted A-type antiferromagnet below $T_N$ = 68 K[12] but its field–cooled magnetization (*M*) shows a cusp around 30 K and *M* changes sign from positive to negative below $T_{comp}$= 9.6 K, where $T_{comp}$ is called compensation point.[13] The occurrence of negative magnetization suggests possible ferrimagnetic interaction in the sample. X-ray magnetic dichroism (XMCD) in the same compound indicates that Mn-3d and Sm-4f moments are antiparallel to each other and anomalous field dependence of magnetocapacitance below and above $T_{comp}$ was also found.[14] Based on neutron diffraction study in $Sm_{1-x}Y_xMnO_3$, Flynn et al.[15] suggested that x = 0.5 composition shows cycloidal ordering of Mn spin below its Neel temperature ($T_N$ = 41 K) but the rare earth moment was suggested not to play significant role in the ordering unlike in $TbMnO_3$. However, the Sr-doped compound $Sm_{0.7}Sr_{0.3}MnO_3$ shows a cusp in the field-cooled magnetization around 30 K below its ferromagnetic Curie temperature.[11] While we can anticipate the increase of $T_c$ with increasing La content in $Sm_{0.7-x}La_xSr_{0.3}MnO_3$, there are no available studies on the resistivity, magnetization, and in particular, the impact of La doping on ferrimagnetic coupling between Sm-4f and Mn-3d moment and magnetocaloric effect in this series. We report structure, electrical resistivity, magnetization and magnetocaloric effect in $Sm_{0.7-x}La_xSr_{0.3}MnO_3$-series ($0 \leq x \leq 0.7$). We also carry out critical exponent analysis and show scaling behavior of the magnetic entropy change for x= 0.6.

## II. Experimental details

Polycrystalline samples of $Sm_{0.7-x}La_xSr_{0.3}MnO_3$ (x= 0 to 0.7) were prepared by the standard solid state synthesis technique. Powder of $Sm_2O_3$, $La_2O_3$, $SrCO_3$ and $Mn_2O_3$ were mixed in appropriate molar fraction. After repeated grinding and calcination at 1000 ºC and 1100 ºC for 12 hours, final sintering was done at 1200 ºC for 24 hours. The samples were characterized by the standard X-ray diffraction technique at room temperature. A commercial vibrating sample magnetometer (Quantum Design Inc., USA) was used for measuring the magnetization. From the isothermal magnetization *M(H)* curves measured at a temperature interval of $\Delta T$= 5 K, $\Delta S_m$ values were estimated using modified Maxwell's equation, $\Delta S_m = \sum_i \left[ M(T_i, H) - M_{i+1}(T_{i+1}, H) \right] \times \Delta H / (T_i - T_{i+1})$, where $M_i$ and $M_{i+1}$ are the magnetization values measured at temperatures $T_i$ and $T_{i+1}$ respectively. Electrical resistivity was measured in the temperature range $T$= 400 K-10 K using a Physical Property Measuring System (PPMS)

## III. Results and Discussion

### A. Structure

Figure 1 shows the X-Ray powder diffraction patterns of the samples x= 0-0.7 of $Sm_{0.7-x}La_xSr_{0.3}MnO_3$. The parent compound x= 0 is orthorhombic (Pnma space group). With increasing amount of La, transition from orthorhombic (Pnma)-to-rhombohedral ($R\bar{3}c$) structure occurs between x values of 0.4 and 0.5. This is evident from the gradual suppression and disappearance of (111) and (021) reflections near 2θ= 25.5°, 34.7° respectively. We have listed the lattice parameters, unit cell volume(V) and bond length(B.L.) in Table 1. It is seen that unit cell volume gradually increases with increasing x from 0 to 0.3 which reflects the larger size of the substituted $La^{3+}$ ion. The structural transition intervenes for x= 0.4.

The tolerance factor which determines the degree of distortion was calculated using the expression $t = \dfrac{<r_A> - r_O}{\sqrt{2}(<r_{Mn}> - r_O)}$ where $<r_A>$, $<r_{Mn}>$ and $r_o$ are the average ionic radii of the A-site, Mn and O ions respectively.[16] The orthorhombic and rhombohedral structures are usually characterized by $t < 1$. We calculated t using average ionic radius with 12 coordination for oxygen and 9 coordination for the A-site cation. The average A-site ionic radius is $<r_A>$ = $(0.7-x)<r_{Sm}^{3+}> + x<r_{La}^{3+}>$. Figure 2 shows the phase diagram, i.e, plot of the ferromagnetic Curie temperature versus the tolerance factor for different x. The La content is shown on the top x-axis. Resistivity and magnetization data were used to obtain this phase diagram as will be discussed in coming sections. With increasing La doping, the tolerance factor increases due to increase in the average ionic radius of the A-site cation which stretches the Mn-O-Mn bond angle and increases Mn-O bond length.

### B. Resistivity

Figure 3(a) shows the temperature dependence of resistivity, $\rho(T)$ measured for x= 0, 0.1, 0.2, 0.3, 0.4, 0.5, 0.55, 0.6, 0.65 and 0.7. The x= 0 sample shows an insulating behavior ($d\rho/dT > 0$). The resistivity below 50 K is beyond the measurement limits of the instrument used. Upon 10% La substitution at the Sm site, the sample shows an insulator to a metal transition with a peak in resistivity occurring at $T_{IM}$ = 90 K. The $T_{IM}$ is very close to the ferromagnetic Curie temperature (see the next paragraph). As x increases, the magnitude of the resistivity peak at $T_{IM}$ decreases and $T_{IM}$ shifts rapidly to high temperature ($T_{IM}$ = 360 K for x= 0.65). Note that the magnitude of $\rho(T)$ at 10 K decreases by over 9 orders of magnitude at 10 K between x= 0 and 0.65. Such a large decrease in resistivity arises from detrapping of charge carries that were localized owing to strong electron-phonon interaction and disorder. The strength of electron phonon interaction decreases with increasing La-content which increases the one electron $e_g$-band width.[17] It is widely accepted that electrical conduction in the high temperature paramagnetic state of manganites is due to thermally activated hopping of polarons. The temperature dependence of the resistivity due to adiabatic polaron hopping follows the relation, $\rho(T) = \rho_0 T \exp\left(\dfrac{E_\rho}{k_B T}\right)$, where $E_\rho$ is the activation energy for small polarons. We have shown the plot of $\ln(\rho/T)$ versus $1/T$ for the x= 0, 0.1 compounds in figure 3(b). Linear fit in the high temperature region gives the activation energy. The inset of figure 3(a) shows the variation of activation energy with composition x. The activation energy decreases with increasing La content as expected.

### C. Magnetization

Figure 4(a) shows the $M(T)$ for x = 0, 0.1, 0.2, 0.3, 0.4, 0.5, 0.55, 0.6, 0.65 and 0.7 measured while cooling under a field of $H$= 1 kOe. Upon cooling, $M(T)$ of x= 0 shows a paramagnetic (PM) to ferromagnetic (FM) transition around $T_c$= 83 K as suggested by a rapid increase of the magnetization. However, the field-cooled $M(T)$ goes through a cusp around $T^*$= 30 K within the long-range FM ordered state and starts decreasing below $T^*$. The $T_c$ increases from 83 K for x= 0 to 375 K for x= 0.7. In the inset (ii) of figure 4(b), we have plotted $T_c$ and $T^*$ versus x. While $T_c$ shows a continuous increase with increasing x, $T^*$ shows a small increase until x= 0.4, after which it increases rapidly and reaches a maximum value of 137 K for x= 0.6 and then decreases. The main panel of figure 4(b) shows the $MH$ loop for the samples of composition x= 0, 0.1, 0.4 and 0.6, indicating that the ground state of the sample is soft ferromagnetic in nature. The magnetization at 5 T (plotted in inset (i) of figure 4(b)) increases from 3.1 $\mu_B$/f.u. for x= 0 to nearly 3.5 $\mu_B$/f.u. for x= 0.1, varies gradually thereafter and increases to 3.6 $\mu_B$/f.u. for x= 0.7. The saturation magnetization of $Sm^{3+}$ moment is expected to be small (= 0.71 $\mu_B$) and it will reduce (increase) the saturation magnetization of Mn sublattice in x= 0 compound by 0.5 $\mu_B$/f.u. if $Sm^{3+}$:4f moment aligns antiparallel (parallel) to the Mn-moment.

Isothermal magnetization ($M$-$H$) curves were measured in both increasing (0-5 T) and decreasing (5-0 T) fields for a temperature interval of 5 K. The coercive field for all the samples were found to be in the range 250-120 Oe and it decreases with increasing La content. Figures 5(a)-5(d) shows the $M$-$H$ curves for x = 0, 0.1, 0.5 and 0.6 at different temperatures. For clarity, we have shown $M(H)$ only for selected temperatures. For the x= 0.6 compound, $M(H)$ is linear in the paramagnetic state and shows a rapid increase at low fields ($H$< 0.5 T) in the ferromagnetic state. However, $M(H)$ at 10 K shows lower values over the 50 K curve in the field range 0-2 T and at higher fields crossover occurs. Such a trend is seen in x= 0.5, 0.1 and 0 samples as well but that the crossover occurs in smaller fields. This is a consequence of the decrease of $M(T)$ below $T^*$ as shown in fig. 4(a). There is no indication of a field-induced metamagnetic transition (i.e., a rapid increase of $M$ above a critical value of $H$) in any of these compounds, which, if present, would have suggested an antiferromagnetic ground state at low temperature. Only the undoped compound (x= 0) shows hysteresis over a larger field range (visible up to 3 T) and the hysteresis is negligible in other compositions above ±300-Oe. The existence of high field hysteresis in x= 0.3 suggests that this compound is not a homogeneous ferromagnet but possibly antiferromagnetic regions coexist with ferromagnetic phase.

### D. Magnetocaloric effect

The magnetic entropy change, $\Delta S_m$, as a function of temperature was evaluated using the Maxwell relation, $\Delta S_m(H,T) = \int_0^H \left(\frac{\partial M}{\partial T}\right)_H dH$. We plot $-\Delta S_m = -[S_m(H)-S_m(0)]$ versus $T$ for x = 0, 0.1, 0.2, 0.3, 0.4, 0.5, 0.6 and 0.7 compounds in figure 6 for (a) $\Delta H$= 1 T and (b) 5 T, respectively. As expected, $-\Delta S_m$ increases with lowering temperature in the paramagnetic state and goes through a maximum value around $T_c$ for each composition. The magnitude of the magnetic entropy change at the peak is nearly the same ($\Delta S_m \sim$ -1.5 J/kg K for $\Delta H$= 1 T) for x= 0.1 to 0.6. The parent compound (x= 0) has slightly lower value (~ 0.9 J/kg K) and the end compound (x= 0.7) has slightly higher value (1.56 J/kg K) compared to other compositions. For $\Delta H$= 5 T, $\Delta S_m$= 4.3±0.2 J/kg K for all x. The x= 0.3 compound shows the maximum $\Delta S_m$ of -4.5 J/kg K for $\Delta H$= 5 T. Much below the ferromagnetic transition, $\Delta S_m$ changes sign from negative to positive at $T= T^*$ for the doped compounds. The positive $\Delta S_m$ is referred as the inverse magnetocaloric effect (inverse MCE) since the magnetic entropy increases upon application of a magnetic field. It means that the sample will be cooled by adiabatic magnetization rather than by adiabatic demagnetization. A horizontal line at $\Delta S_m$= 0 separates the normal MCE and the inverse MCE. The magnitude inverse MCE is much smaller than the peak value the normal MCE at $T_c$ because the inverse MCE occurs within the long range ferromagnetic state.

In addition to a large $\Delta S_m$ value, a large refrigerant capacity (RC) is highly preferred for practical application. We compute RC using the equation, $RC = \int_{T_1}^{T_2} \Delta S_m(T) dT$, where $T_1$ and $T_2$ are the temperatures corresponding to the half-maximum of the $\Delta S_m(T)$ peak around $T_c$. We plot the peak value of $-\Delta S_m$, RC for $\Delta H$= 1, 3 and 5 T and $T_c$ as a function of composition in fig. 7(a)-(c). The RC for $\Delta H$= 5 T is highest ≈ 255 J/kg K in x= 0, which has the lowest $T_c$. The RC decreases with increasing x. However, RC is still significant, for example, RC= 181 J/kg K for $\Delta H$= 5 T in x= 0.5 having $T_c$= 307 K. The RC and $\Delta S_m$ for this composition for $\Delta H$= 1 T (2 T) are 35.85 (73.80) J/kg and 1.28 (2.24) J/Kg K respectively. The magnetocaloric parameters of all the samples are listed in Table 2. We compare the magnetocaloric parameters of samples with $T_c$ around room temperature reported in literature[18,19,20,21] along with our samples in Table 3.

Figure 7(d) is a scaling plot of $\Delta S_m/\Delta S^{max}$ (normalized $\Delta S_m$) versus $T/T_c$ which is helpful to quickly identify the compound with the highest RC among several compounds. With increasing x, the width of the plot decreases until x= 0.3 and remains almost constant for all higher x values. This indicates that the refrigerant capacity decreases initially with increasing x and remains constant thereupon which is also evident from figure 7(b).

### E. Critical behavior near the phase transition

From the isothermal magnetization data presented in fig. 5(d) we plot $\mu_0 H/M$ versus $M^2$ (Arrott plots) for x= 0.6 composition in Figure 8. Arrott plot is generally used to verify second order paramagnetic to ferromagnetic transition and also to extract critical parameters associated with the phase transition.[22] The linearity at high fields and positive slope exhibited by the Arrott plots above $T_c$ confirm the second order nature of the FM to PM transition.[23] Using these plots, we determine the Curie temperature and the critical exponents in the vicinity of the phase transition temperature. By extrapolating the Arrott plots to $H/M = 0$ for $T < T_c$ and $M^2 = 0$ for $T > T_c$, the spontaneous magnetization $M_s(T)$ and the inverse initial susceptibility $\chi_0^{-1}(T)$ were calculated. The exponents $\beta$ and $\gamma$ were evaluated by fitting equations (1) and (2) to the plots of $M_s(T)$ and $\chi_0^{-1}(T)$ (Figure 9 (a)), respectively. The critical isotherm exponent, $\delta$, is evaluated from the Widom scaling relation $\delta = 1 + \frac{\gamma}{\beta}$.

$$M_s(T) = M_0 |\varepsilon|^\beta, \quad \varepsilon < 0, T < T_c \quad (1)$$

$$\chi_0^{-1}(T) = (h_0/M_0)\varepsilon^\gamma, \quad \varepsilon > 0, T > T_c \quad (2)$$

$$M = DH^{1/\delta}, \varepsilon = 0, \quad T = T_c \quad (3)$$

where $\varepsilon = (T - T_c)/T_c$ is the reduced temperature and $M_0$, $(h_0/M_0)$ and $D$ are the critical amplitudes.[24] From the analysis, we obtained $T_c = 352$ K, $\beta = 0.43$ and $\gamma = 1.03$. The exponent $\delta$ is obtained from the Widom scaling relation as 3.395. These obtained values are close to that predicted by the mean field model ($\beta = 0.5$, $\gamma = 1$ and $\delta = 3$)[25].

According to the mean field model, $\Delta S_m$ and $M^2$ for small $M$ values are related by the equation[26]

$$-\Delta S_m(M_s(T)) = \frac{1}{2C_M}\left[M(T)^2 - M_s(T)^2\right] \quad (4)$$

where $C_M$ is the Curie constant. Inset of figure 8 shows isothermal plots of magnetic entropy ($-\Delta S_m$) vs. $M^2$ which shows a linear behavior with a nearly constant slope in agreement with equation (4). $M_s(T)$ values at different temperatures are obtained by linear extrapolation of these curves. The close agreement of these values with that obtained from Arrott plot is shown in Figure 9(b). These results indicate that the mean field theory clearly describes the ferromagnetic phase transition in this compound.

Recently, V. Franco *et al*. suggested that magnetic entropy change under different magnetic fields for a second order PM-FM transition collapses into a master curve when proper scaling of temperature is introduced.[27] We would like to verify their idea in our x= 0.6 compound. We plotted the normalized $\Delta S_m$ (i.e., $\Delta S_m/\Delta S_{max}$) against the reduced temperature $\theta = \frac{T-T_P}{T_r-T_P}$ where $T_P$ is the temperature at which $\Delta S_m$ peaks and $T_r$ is the reference temperature corresponding to $\Delta S_m = \Delta S_{max}/2$. Figure 10 shows the universal behavior for x= 0.6 sample thus evidencing the second order nature of the PM-FM transition in this compound.

### F. Discussion

The most important results in our study are (1) Composition driven insulator to metal transition, (2) Observation of a peak in the magnetization at the temperature $T^*$ well within the ferromagnetic region in all but the x= 0.7 compound, (3) Occurrence of the inverse magnetocaloric effect below $T^*$ and nearly constant value of $\Delta S_m$ for a wide range of composition. The observation, i.e., composition driven insulator-metal transition can be ascribed to the increase in one electron bandwidth due to increasing the tolerance factor. Since the average ionic radius of the $La^{3+}$ is larger than $Sm^{3+}$, the Mn-O-Mn bond angle and also Mn-O bond length are supposed to increase with increasing $La^{3+}$ content. In $R_{0.7}Sr_{0.3}MnO_3$ manganites, compounds with R= La, Pr and Nd are ferromagnetic metals whereas R= Gd is a spin or cluster glass insulator.[28] From our results, it appears that R= Sm seems to be the border line case which is a ferromagnetic insulator. However, x= 0 is not a homogeneous ferromagnet as suggested by hysteresis in *M-H* curves up to 3 T (Fig. 3). In this composition, the antiferromagnetic superexchange interaction between $t_{2g}^3$ spins competes with ferromagnetic double exchange interaction and localizes $e_g$ holes below $T_c$. As the $e_g$ electron band width increases with La doping, the electron-phonon coupling decreases and the ferromagnetic state becomes more homogenous. As a result, $T_c$ increases and peak value of the ρ(*T*) decreases with increasing x. Interestingly, all the compounds investigated in this series show second-order PM-FM transition in magnetization and no hysteresis is observed in resistivity between cooling and warming.

A peculiar behavior found in this series is the occurrence of a cusp in the field-cooled *M*(*T*) much below $T_c$. Similar cusp in *M*(*T*) below $T_c$ in ac susceptibility was observed in earlier $Sm_{1-x}Sr_xMnO_3$ series for x= 0.3-0.5.[11,12,29,30,31] Borges *et al.*[18] interpreted that the cusp is due to increase of coercive field with lowering temperature. However, the coercive field of x= 0.6 sample is only 120 Oe even at 10 K, which indicates that the origin of the low temperature anomaly is not due to increase in coercive field. We suggest that the cusp in $Sm_{0.7}Sr_{0.3}MnO_3$ is related to the ordering of Sm-$4f^5$ moment antiparallel to the Mn-moments and change in the easy axis of magnetization (spin reorientation transition, SRT) of the Mn-sublattice occurring concurrently. The rare earth moment ordering is induced by the exchange field of the Mn-sublattice as like in $SmMnO_3$.[13] Due to strong neutron absorption by Sm, investigation of the Sm moment ordering in La doped compounds by neutron diffraction is difficult. Element specific techniques such as XMCD will be useful to pinpoint the ordering of the Sm moment in our compounds. Since $T_c$ increases with $La^{3+}$ content, the exchange field felt by the Sm-4f site also increases. Hence, Sm-4f moments can be polarized antiparallel to the Mn-sublattice at a higher temperature. The competition between magnetocrystalline anisotropy of the Mn lattice and single ion anisotropy of $Sm^{3+}$ may lead to spin reorientation transition in the Mn-sublattice, which occurs at a higher temperature for larger x. However, when La content increases above 0.6 or Sm content decreases by 86 %, the SRT shifts down in temperature. The SRT is not an uncommon phenomenon in rare (R) earth based $RFeO_3$ (R = Tb, Sm. Ho, Er etc.) insulating orthoferrites.[32,33] In orthoferrites, Fe ions order antiferromagentically at high temperatures (> 500 K) and canting of Fe spins due to Dzyaloshinskii-Moriya (DM) interaction gives rise to a weak ferromagnetic component in the antiferromagnetic state. Competition between single ion anisotropy of R and anisotropic dipolar interactions leads to SRT at a lower temperature, which can be second order or first order.[34] Unlike the orthoferrites, our compounds are ferromagnetic and metallic. The exact origin of the SRT in our compounds needs a detailed study. Nevertheless, we would like to bring to the notice of readers that ordering of Sm-4f moments induced by exchange fields of transition metal ion lattice occurs in other Sm-based compounds. The canted antiferromagnet $SmFeO_3$ shows a cusp in *a*-axis magnetization around 130 K much below the Neel temperature of Fe-sublattice.[35] The ferromagnetic pyrochlore $Sm_2Mo_2O_7$ shows a cusp around 35-40 K below the ferromagnetic ordering of Mo:$t_{2g}^2$ spins due to polarization of $Sm^{3+}$ antiparallel to the ferromagnetic moment Mo sublattice.[36] Ferrimagnetic coupling between $Sm^{3+}$ and $Ti^{4+}$ also seems to be prevalent in the titanate $La_{1-x}Sm_xTiO_3$ for certain value of x.[37] A strong exchange interaction between Sm and Fe sublattices and existence of induced ordering of Sm at temperature as high as 110 K, which coincides with ordering temperature of Fe moments, were also noted in the pnictide SmFeAsO.[38]

Inverse MCE is the general property of a material whose magnetization decreases with decreasing temperature. However, the origin of decrease of magnetization will be different in different materials. In Heusler alloy such as $Ni_{50}Mn_{34}In_{16}$, the inverse MCE is associated with high moment austentite to low moment martensite diffusionless structural phase transition.[39,40] In manganites such as $NdBaMn_2O_6$[41], $La_{0.125}Ca_{0.875}MnO_3$[8],

Pr$_{0.5}$Sr$_{0.5}$MnO$_3$[42], Pr$_{0.46}$Sr$_{0.54}$MnO$_3$[43] the inverse MCE is related to antiferromagnetic transition from either ferromagnetic or paramagnetic phase upon cooling. Inverse MCE in La$_{0.7}$Sr$_{0.3}$MnO$_3$/SrRuO$_3$ superlattices[44] is associated with AFM coupling between FM layers. Inverse MCE also occurs due to spin-reorientation transition (examples NdCo$_5$[45], Nd$_2$Co$_7$)[46] or ferrimagnetic interaction between rare earth sublattices (example Gd$_{1-x}$Pr$_x$Al$_2$).[47,] Oscillatory sign of magnetic entropy is predicted even in diamagnets due to field-induced changes in the density of states at the Fermi level.[48] However, inverse MCE in our Sm-doped compounds is due to induced polarization of Sm-4f moment and accompanying spin reorientation transition in the Mn-sublattice. Coexistence of both normal and inverse magnetocaloric effects in a single sample is interesting for applications because cooling can be achieved both by adiabatic magnetization as well as demagnetization and also increases the temperature range of operation for magnetic refrigeration. When used in composite form, the magnetic entropy change can apparently remain constant over a temperature range of 200 K, which is very much desirable for practical application. The fact that the series has an almost constant $\Delta S_m$ value with tunable $T_c$ makes these compounds interesting for application over a wide temperature range.

**Conclusion**

In summary, we have presented electrical resistivity, magnetization and magnetic entropy change in Sm$_{1-x}$La$_x$Sr$_{0.3}$MnO$_3$. It is found that the insulating Sm$_{0.7}$Sr$_{0.3}$MnO$_3$ transforms into ferromagnetic metal for x= 0.1 and the ferromagnetic Currie temperature increases with increasing La content. All but x= 0.7 (La only) compounds showed unusual maximum in the field-cooled magnetization at $T= T^*$, well within the ferromagnetic state. We attribute this anomaly to 4f-3d antiferromagnetic interaction between the rare earth and transition metal sublattices. The sign of magnetic entropy (-$\Delta S_m$) was found to be negative below $T^*$ (inverse MCE) and positive above $T^*$. The magnetic entropy change at the Curie temperature is nearly constant ($-\Delta S_m$= 4.3±0.2 J/kg K for $\Delta H$= 5 T) for all the compositions and the refrigerant capacity is ~250 J/kg. The tunability of $T_c$ over a wide temperature range with high values of magnetic entropy change and refrigerant capacity makes this system a potential candidate for efficient refrigeration over remarkably broad operating temperatures. Critical exponents associated with the high temperature paramagnetic to ferromagnetic transition were analyzed for x= 0.6 composition and the obtained values are in close agreement with mean-field model.

**Acknowledgement**

R.M acknowledges the National Research Foundation, Singapore (Grant no. NRF-CRP-G-2007-12) for supporting this work.

**Figure Captions**

**Figure 1.** X-ray diffraction pattern of $Sm_{0.7-x}La_xSr_{0.3}MnO_3$ (x= 0 - 0.7) compounds.

**Figure 2.** Phase diagram for $Sm_{0.7-x}La_xSr_{0.3}MnO_3$ (x= 0 - 0.7) as a function of tolerance factor *t* and composition x. $T_c$ and $T^*$ correspond to ferromagnetic Curie temperature of the Mn-sublattice and occurrence of a cusp in the field-cooled magnetization measured at H = 1 kOe.

**Figure 3.** (a) Temperature dependence of the resistivity $\rho(T)$ of x= 0 - 0.6 in zero magnetic field. The inset shows the activation energy for polaron hopping with varying x. (b) $\ln(\rho/T)$ versus 1/T plots with linear fit at high temperatures for x= 0 and 0.1.

**Figure 4.** (a) Field-cooled magnetization *M(T)* for x= 0 - 0.7 under *H*= 1 kOe. (b) *M(H)* at 10K for x= 0, 0.1, 0.4, 0.6. The inset (i) shows dependence of magnetization at 5 T at 10 K on composition. The inset (ii) shows the $T_c$ and $T^*$ as a function of composition x.

**Figure 5.** M(H) plots at selected temperatures for (a) x= 0, (b) 0.1, (c) 0.5 and (d) 0.6 compounds

**Figure 6.** Temperature dependence of the magnetic entropy ($\Delta S_m$) obtained from *M(H)* data at (a) $\Delta H = 1T$ and (b) 5T for x= 0 to 0.7. Inset shows the variation of maximum magnetic entropy with magnetic field for all compositions.

**Figure 7.** Values of (a) $\Delta S_m$ at $T_c$ and (b) refrigerant capacity (RC) for $\Delta H$= 1, 2 and 5 T and (c) $T_c$ as a function of composition x. (d) Normalized $\Delta S_m$ versus $T/T_c$ for different x.

**Figure 8.** Arrott plots ($\mu_0 H/M$ vs. $M^2$) of isothermal magnetization. Inset shows isothermal (-$\Delta S_m$) vs $M^2$ curve of $Sm_{0.1}La_{0.6}Sr_{0.3}MnO_3$

**Figure 9.** (a) Spontaneous magnetization and inverse initial susceptibility deduced by extrapolating Arrot plot ($\mu_0 H/M$ vs. $M^2$) to $\mu_0 H = 0$ and $M^2 = 0$, respectively. Solid lines are best fits to Eqs. (1) and (2) in the text. (b) Spontaneous magnetization of $Sm_{0.1}La_{0.6}Sr_{0.3}MnO_3$ estimated from (-$\Delta S_m$) vs $M^2$ curve and Arrott plots.

**Figure 10.** Normalised $\Delta S_m$ versus normalized temperature $\theta$ for different applied magnetic fields for $Sm_{0.1}La_{0.6}Sr_{0.3}MnO_3$.

**Reference**

[1] A. Gschneidner Jr, V.K. Pecharsky, and A.O. Tsokol, Rep. Prog. Phys. **68,** 1479 (2005); N.A. de Oliveira and P.J. von Ranke, Phys. Rep. **489**, 89 (2010).

[2] E. Brück, Journal of Physics D: Applied Physics **38,** R381 (2005)

[3] B.G. Shen, J.R. Sun, F.X. Hu, H.W. Zhang, and Z.H. Cheng, Adv. Mat. **21**, 4545 (2009)

[4] D.T. Cam Thanh, E. Bruck, O. Tegus, J.C.P. Klaasse, T.J. Gortenmulder, and K.H.J. Buschow, J.Appl. Phys. **99**, 08Q107 (2006).

[5] See, for a review, M.H. Phan and S.C. Yu, J. Magn. Magn. Mater. **308**, 325 (2007)

[6] S. Jeppesen, S. Linderoth, N. Pryds, L.T. Kuhn and J.B. Jensen, Rev. Sci. Instrum. **79**, 083901, (2008)

[7] A.N. Ulyanov, J.S. Kim, G.M. Shin, Y.M. Kang and S.I. Yoo, J. Phys. D: Appl. Phys. **40,** 123 (2007)

[8] A. Biswas, T. Samanta, S. Banerjee, and I. Das, Appl. Phys. Lett. **94**, 233109 (2009)

[9] Y. D. Zhang, P.J. Lampen, T.L. Phan, S.C. Yu, H. Srikanth, and M-H.Phan, J. Appl. Phys. **111**, 063918 (2012)

[10] C.R.H. Bahl, R. Bjørk, A. Smith, and K.K. Nielsen, J. Magn. Magn. Mater. **324**, 564 (2012); C.R.H. Bahl, D. Velazquez, K.K. Nielsen, K. Engelbrecht, K.B. Andersen, R. Bulatova, and N. Pryds, Appl. Phys. Lett. **100**, 121905 (2012)

[11] A. Rebello and R. Mahendiran, Appl. Phys. Lett. **93,** 232501 (2008)

[12] V.Y. Ivanov, A.A. Mukhin, A.S. Prokhorov, and A.M. Balbashov, Phys. Stat. Sol. B **236**, 445 (2003)

[13] J.-G. Cheng, J.-S. Zhou, J.B. Goodenough, Y. T. Su, Y. Sui, and Y. Ren, Phys. Rev. B **84**, 104415 (2011)

[14] J-S. Jung, A. Iyama, H. Nakamura, M. Mizumaki, N. Kawamura, Y. Wakabayashi, and T. Kimura, Phys. Rev. B **82**, 212403 (2010);

[15] D. O'Flynn, C.V. Tomy, M.R. Lees and G. Balakrishnan, Phys. Rev.B **83**, 174426 (2011)

[16] J.M.D. Coey, M. Viret and S.V. Molnar, Adv. Phys. **48**, 167 (1999)

[17] W. Archibald, J.-S. Zhou, and J.B. Goodenough, Phys. Rev. B **53**, 14445 (1996)

[18] Y.D. Zhang, P.J. Lampen, T.L. Phan, S.C. Yu, H. Srikanth, and M.H. Phan, Journal of Applied Physics **111,** 063918 (2012)

[19] M.A. Choudhury, S. Akhter, D.L. Minh, N.D. Tho, N. Chau, J. Magn. Magn. Mater. **1295**, 272 (2004)

[20] N. Chau, P.Q. Niem, H.N. Nhat, N.H. Luong, N.D. Tho, Physica B **214**, 327 (2003)

[21] W. Chen, L.Y. Nie, W. Zhong, Y.J. Shi, J.J. Hu, A.J. Li, Y.W. Du, J. Alloys Compd. **23**, 395 (2005)

[22] M. Koubaa, W. Cheikhrouhou-Koubaa , A. Cheikhrouhou, and L. Ranno, Physica B **403,** 4012 (2008)

[23] B.K. Banerjee, Phys. Lett. **12,** 16 (1964)

[24] A. Arrott, Phys. Rev. **108,** 1394 (1957)

[25] A. Arrott and J. E. Noakes, Phys. Rev. Lett. **19,** 786 (1967)

[26] S. Kallel, N. Kallel, O. Peña, and M. Oumezzine, J. Alloys Comp. **504,** 12 (2010)

[27] V. Franco, J.S. Blazquez, and A. Conde, Appl. Phys. Lett.**89,** 222512 (2006); C.M. Bonilla, J. Herrero-Albillos, F. Bartolome , L.M. Garcia, M. Parra-Borderias and C. Franco, Phys. Rev. B **81**, 22424 (2011)

[28] J.P. Zhou, J.T. McDevitt, J.S. Zhou, H.Q. Yin, J.B. Goodenough, Y. Gim and Q.X. Jia, Appl. Phys. Lett. **75**, 1146 (1996)

[29] R.P. Borges, F. Ott, R.M. Thomas, V. Skumryev,  J.M.D. Coey, J.I. Arnaudas, and L. Ranno, Phys. Rev. B **60**, 12847 (1999)

[30] M. Eglimez, K.H. Chow, J. Jung, and Z. Salman, Appl. Phys. Lett. **90**, 162508 (2007); *ibid* **92**, 132505 (2008)

[31] N.A. Babushkina, E.A. Chistotina, O. Yu. Gorbenko, A.R. Kaul, D.I. Khomskii, and K.I. Kugel, Phys. Rev. B **67**, 100410 (2003)

[32] G. Gorodetsky and L. M. Levinson, Solid State Commun.**7**, 67 (1969); L. M. Livingston and S. Shtrikman, Solid State Commun. **8**, 209 (1970):L. M. Livingston, M. Luban, S. Shtrikann, Phys. Rev. B **187**, 715 (1969)

[33] R. D. Pierce, R. Wolfe, and L. G. Van Uitert, J. Appl. Phys. 40, 1241 (1969). Y. Sundarayya, P. Mandal, A. Sundaresan and C. N. R Rao, J. Phys. :Condens. Matter. **23**, 436001 (2011)

[34] Beom Hyun Kim and B. I. Min, Phys. Rev. B **77**, 094429 (2008) and references therein.

[35] L.G. Marshall, J.-G. Cheng, J.-S. Zhou, J. B. Goodenough, J.-Q. Yan, and D. G. Mandrus, Phys. Rev. B **86**, 064417 (2012)

[36] N. Ali, P. Hill, S. Labord and J.E. Greedan, J. Solid State Chem. **83**, 178 (1989); Surjeet Singh, R. Suryanarayanan, R. Tackett, G. Lawes, A.K. Sood, P. Berthet, and A. Revcolevschi, Phys. Rev. B **77**, 020406R (2008)

[37] K. Yoshii and A. Nakamura, J. Solid State Chem. **133**, 584 (1997)

[38] S. Nandi, Y. Su, Y. Xiao *et al.* Phys. Rev. B **84**, 054419 (2011)

[39] V. K. Sharma, M.K. Chattopadhyay and S.B. Roy, J. Phys. D. Appl. Phys. **40**, 1869(2007)

[40] T. Krenke, E. Duman, M. Acet, E.F. Wassermann, X. Moya, L. Manosa, and A. Planes, Nature Maters. **4**, 450 (2005)

[41] Q. Zhang, F. Guillou, A. Wahl, Y. Bŕard, and V. Hardy, Appl. Phys. Lett. **96**, 242506(2010);A. M. Aliev, A.G. Gamsatov, V.S. Kalitak and A.R. Kaul, Solid State Comm.**151**, 1820 (2011)

[42] N.S. Bingham, M.H. Phan, H. Srikanth, M.A. Torija, and C. Leighton, J. Appl. Phys. **106**, 023909 (2009)

[43] V. B. Naik, S. K. Barik, R. Mahendiran, and B. Raveau, Appl. Phys. Lett. **98**, 112506 (2011)

[44] S. Thota, Q. Zhang, F. Guillou, U. Lüders, N. Barrier, W. Prellier, A. Wahl, and P. Padhan, Appl. Phys. Lett. **97**, 112506 (2010)

[45] S.A. Nikitin, K.P. Skokov, Yu.S. Koshkid'ko, Uy. G. Pastushenkov, and T.I. Ivanova, Phys. Rev. Lett. **105**, 137205 (2010).

[46] M. Ilyn, M.I. Bartashevich, A.V. Andreev, E.A. Tereshina, V. Zhukova, A. Zhukov, and J. Gonzalez, J. Appl. Phys. **109**, 083932 (2011)

[47] V.S.R. de Sousa, A.Magnus G.Carvalho, E.J.R.Plaza, B.P.Alho, J.C.G.Tedesco, A.A. Coelho, N.A.deOliveira, P.J.von Ranke, J. Magn. Magn. Mater. 323, 794 (2011).

[48] M.R. Reiss, Appl. Phys. Lett. **99**, 052511 (2011)

| Parameter | Orthorhombic - Pnma | | | | | Rhombohedral - R$\bar{3}$c | | |
|---|---|---|---|---|---|---|---|---|
| | x = 0 | x = 0.1 | x = 0.2 | x = 0.3 | x = 0.4 | x = 0.5 | x = 0.6 | x = 0.7 |
| a (Å) | 5.436 | 5.460 | 5.470 | 5.479 | 5.484 | 5.489 | 5.497 | 5.499 |
| b (Å) | 7.68 | 7.709 | 7.724 | 7.745 | 7.75 | | | |
| c (Å) | 5.469 | 5.47 | 5.4705 | 5.465 | 5.49 | 13.351 | 13.355 | 13.358 |
| V (Å$^3$) | 228.36 | 230.26 | 231.13 | 231.944 | 232.479 | 349.06 | 349.48 | 349.82 |
| B. L. (Å) | 1.9301 | 1.9507 | 1.9519 | 1.9534 | 1.9547 | 1.95661 | 1.95759 | 1.95824 |

**TABLE 1**. Lattice parameters calculated from XRD scan for different compositions

| x | $T_c$ (K) | $E_\rho$ (meV) | $\Delta H = 1$ T | | $\Delta H = 2$ T | | $\Delta H = 5$ T | |
|---|---|---|---|---|---|---|---|---|
| | | | $\Delta S_m$ (J/kg.K) | RC (J/kg) | $\Delta S_m$ (J/kg.K) | RC (J/kg) | $\Delta S_m$ (J/kg.K) | RC (J/kg) |
| 0 | 83 | 150.1 | 0.884 | 44.41 | 1.51 | 97.25 | 3.21 | 253.43 |
| 0.1 | 95 | 147.6 | 1.381 | 43.48 | 2.29 | 102.56 | 4.14 | 257.06 |
| 0.2 | 123 | 146.6 | 1.299 | 43.7 | 2.38 | 95.29 | 4.45 | 255.17 |
| 0.3 | 195 | 137.0 | 1.34 | 48.9 | 2.44 | 85.96 | 4.51 | 217.51 |
| 0.4 | 256 | 111.6 | 1.283 | 32.38 | 2.318 | 73.04 | 4.41 | 199.01 |
| 0.5 | 297 | 78.6 | 1.28 | 35.33 | 2.24 | 73.80 | 4.26 | 195.56 |
| 0.6 | 338 | 66.2 | 1.312 | 33.44 | 2.25 | 63.50 | 4.29 | 180.62 |
| 0.7 | 375 | - | 1.44 | 32 | 2.49 | 65.60 | 4.51 | 179.50 |

**TABLE 2**. Curie temperature, Activation energy, Maximum entropy change and RC for different compositions

| Material | $\Delta H$ (T) | $T_c$ (K) | $\Delta S^{max}$ (J/kg.K) | RC (J/kg) | Reference |
|---|---|---|---|---|---|
| $La_{0.68}Pr_{0.02}Sr_{0.3}MnO_3$ | 1.5 | 354 | 2.79 | 44 | 18 |
| $La_{0.7}Sr_{0.3}Mn_{0.98}Ni_{0.02}O_3$ | 1.35 | 325 | 3.54 | 71 | 19 |
| $La_{0.7}Sr_{0.3}Mn_{0.95}Cu_{0.05}O_3$ | 1.35 | 350 | 1.96 | 39 | 20 |
| $La_{0.65}Nd_{0.05}Ba_{0.3}MnO_3$ | 1 | 325 | 1.57 | 24 | 21 |
| $La_{0.5}Sm_{0.2}Sr_{0.3}MnO_3$ | 1.5 | 297 | 1.94 | 73 | This work |
| $La_{0.6}Sm_{0.1}Sr_{0.3}MnO_3$ | 1.5 | 338 | 1.80 | 51 | This work |

**TABLE 3**. Magnetocaloric parameters at the Curie temperature $T_C$ for different manganites.

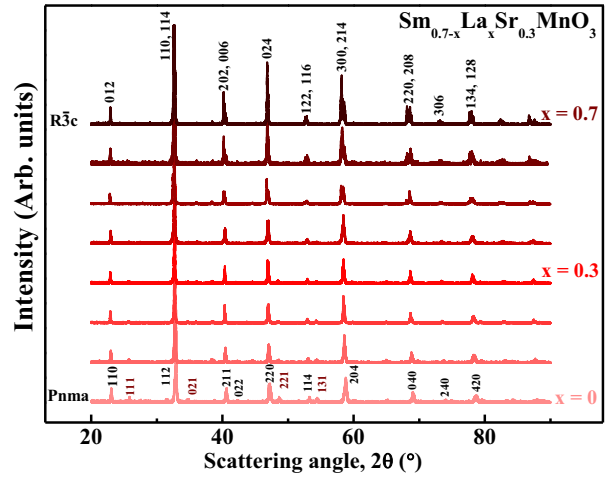

Figure 1.

M. Aparnadevi *et al*

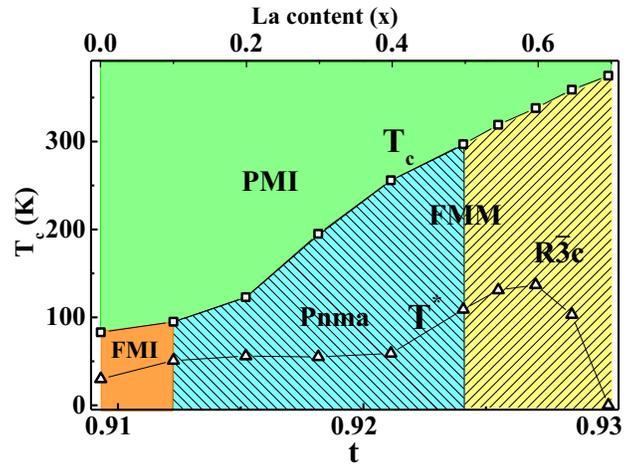

Figure 2.

M. Aparnadevi *et al*

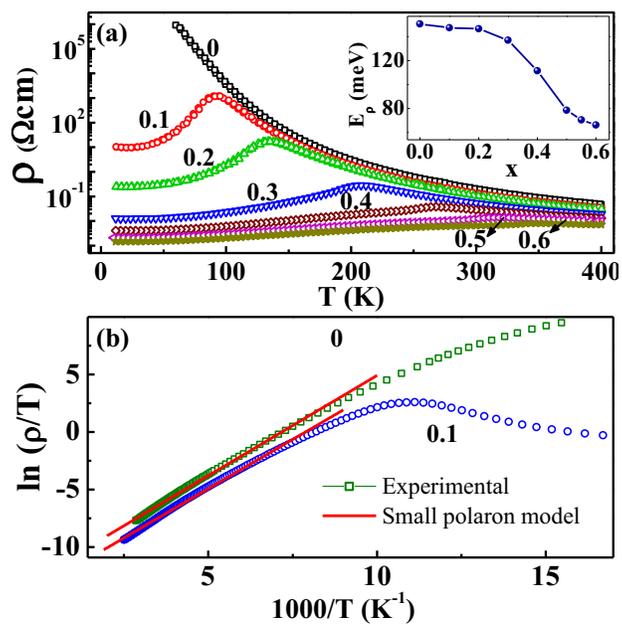

Figure 3

M. Aparnadevi *et al*

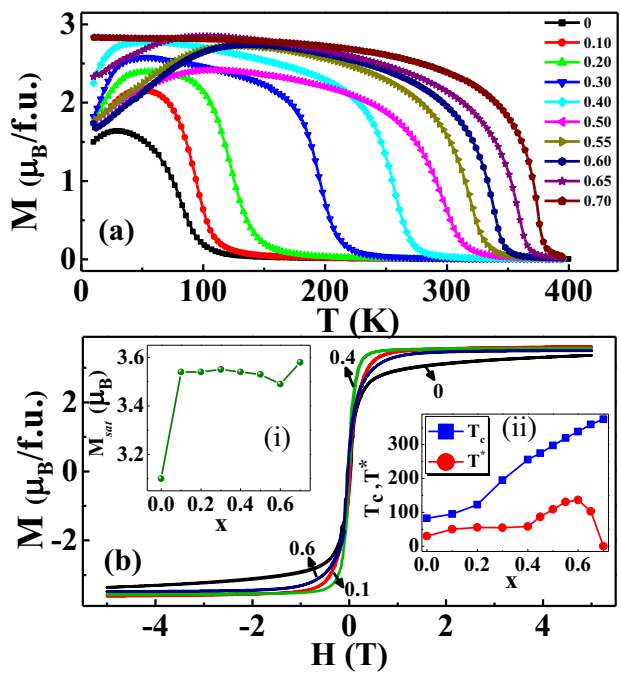

Figure 4

M. Aparnadevi *et al*

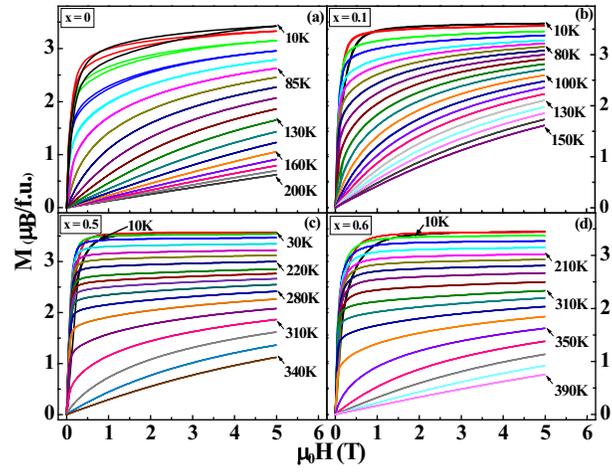

Figure 5

M. Aparnadevi *et al*

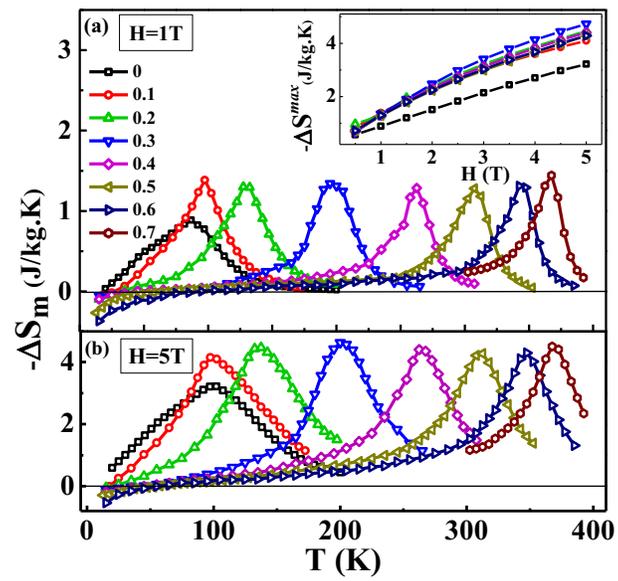

Figure 6

M. Aparnadevi *et al*

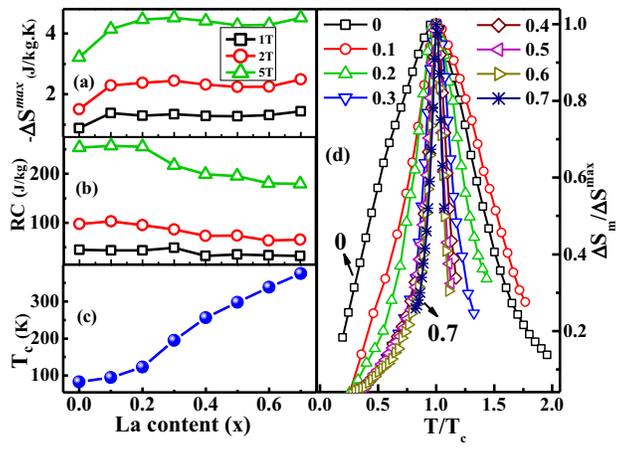

Figure 7

M. Aparnadevi *et al*

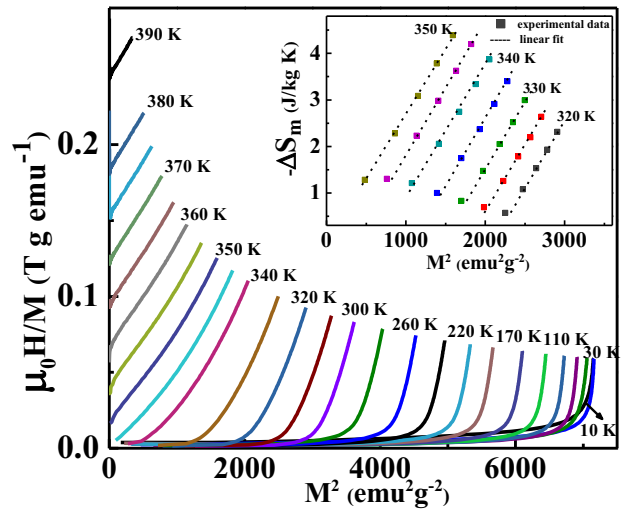

Figure 8

M. Aparnadevi *et al*

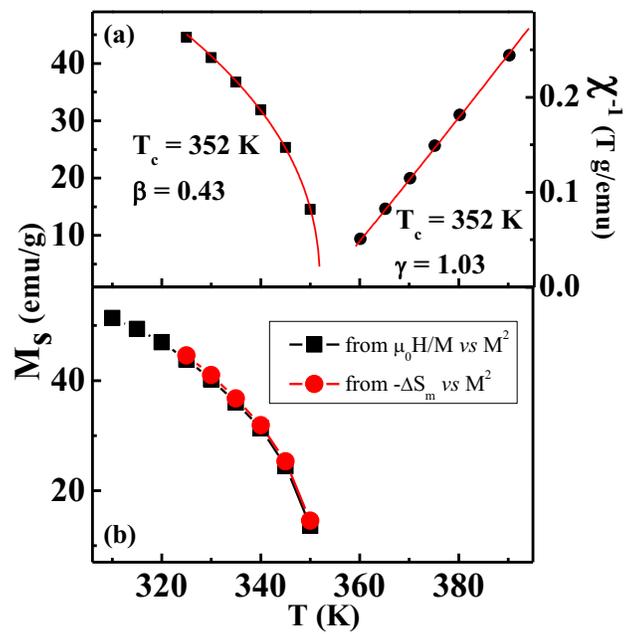

Figure 9

M. Aparnadevi *et al*

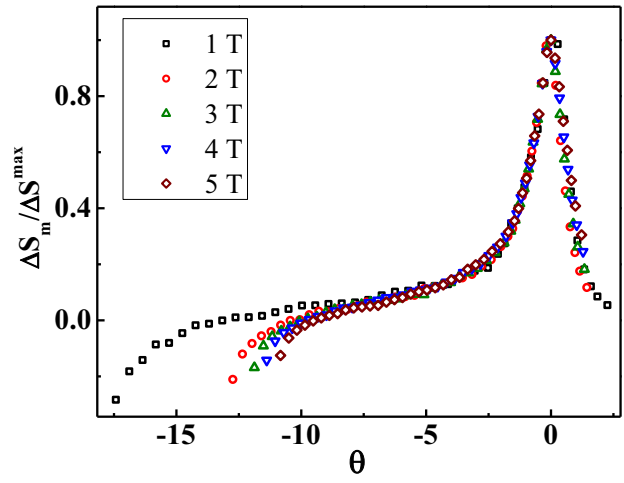

Figure 10

M. Aparnadevi *et al*